\documentclass[10pt,letterpaper,twocolumn]{article} %% two column, final layout

\usepackage{ol2}
\usepackage[draft]{hyperref}
\usepackage{amsmath}

\begin{document}

\twocolumn[ %% activate for two-column option

\title{Direct measurement of the Wigner time-delay\\ for the scattering of light by
a single atom}

\author{R.~Bourgain, J.~Pellegrino, S.~Jennewein, Y.R.P.~Sortais$^*$ and A.~Browaeys}

\address{
Laboratoire Charles Fabry, Institut d'Optique, CNRS, Univ Paris Sud, \\
2 Avenue Augustin Fresnel, 91127 Palaiseau cedex, France\\
$^*$Corresponding author: yvan.sortais@institutoptique.fr
}

\begin{abstract}We have implemented the gedanken experiment of an individual atom
scattering a wave packet of near-resonant light, and measured the associated
Wigner time-delay as a function of the frequency of the light. In our apparatus
the atom behaves as a two-level system and we have found delays as large as $42$
nanoseconds at resonance, limited by the lifetime of the excited state. This delay
is an important parameter in the problem of collective near-resonant scattering
by an ensemble of interacting particles, which is encountered in many areas of physics.
\end{abstract}


 ] %% activate for two-column option

The scattering of an incident wave, whether classical or quantum, upon a single particle
is not an instantaneous process. In 1955, E.P.~Wigner showed that the time-delay
associated to the elastic scattering of a wave upon a scatterer is the derivative of the
phase shift aquired by the incident wave with respect to its
energy~\cite{Wigner1955}. Later, F.T.~Smith pointed out that this so-called Wigner
delay is the lifetime of a resonant state excited during the scattering~\cite{Smith1960},
therefore being largest at resonance. Since its derivation, the Wigner delay has been used
as an important parameter in the problem of near-resonant scattering in dense media,
as it governs the transport of energy~\cite{Lagendijk1996,Nussenzveig2002,Labeyrie2003,MullerPRA2005%,PierratPRA2009
,PierratPRA2010}.

Scattering processes are common in many areas of physics, and therefore many systems
are candidates for the measurement of this delay. However, in most systems it
is expected to be very short, explaining why experimental demonstrations are
scarce and have required involved techniques. In 1976, a delay in the $10^{-20}$ s
range was measured in the elastic near-resonant scattering of protons on a target of
carbon using interferences in the bremsstrahlung radiation~\cite{Maroni1976}.
The advent of ultra-short pulse laser-based metrology made it possible to measure
time-delays associated to the scattering of light by condensed matter systems or
by atomic vapors. Femtosecond laser techniques, for instance, allowed to measure delays
of a few femtoseconds associated to the bouncing of light off a metallic surface in the
vicinity of a plasmon resonance~\cite{Chauvat2000}. More recently, attosecond
metrology led to the measurement of delays in the $10-100$~attoseconds range in
the photo-emission from a surface~\cite{Cavalieri2007} or from a
vapor~\cite{Schultze2010,Klunder2011}. Yet, the gedanken experiment imagined
initially by Wigner has never been realized, i.e. the direct measurement of the
time-delay induced by a single scatterer upon an incident wave. Here, we do so
by sending a gaussian wavepacket of near-resonant light on an isolated individual
atom and by scanning the frequency of the light across an atomic resonance.

\begin{figure}
\centerline{\includegraphics[width=8cm]{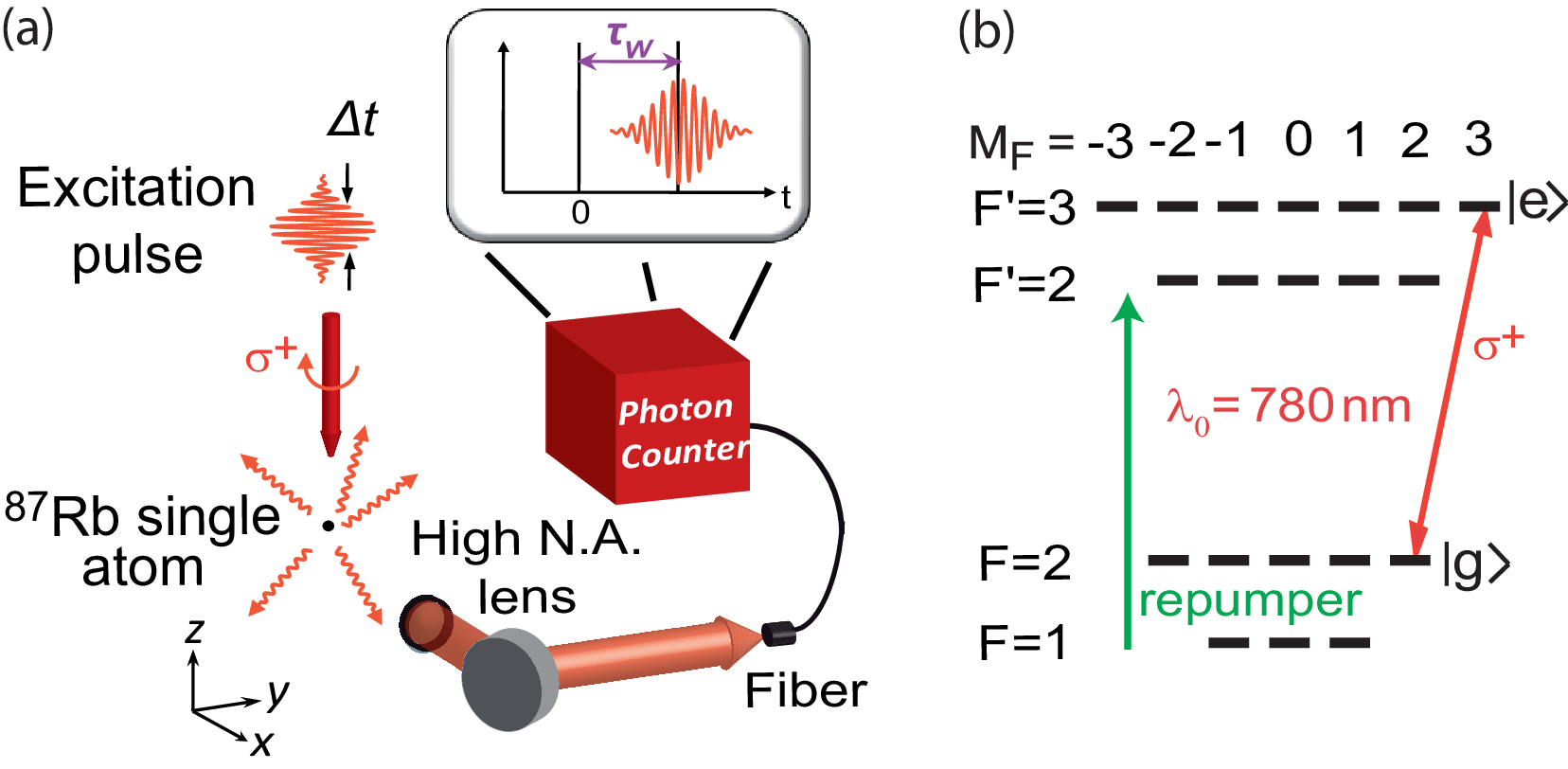}}
\caption{(a) Implementation of the Wigner gedanken experiment : we excite a single atom by a pulse of light and measure the arrival time of the scattered pulse on a photon counter. We use the large numerical aperture lens to isolate a single atom with a tightly focused laser beam (not shown), and collect the scattered photons efficiently. (b) Atomic levels used in the experiment.}\label{fig1_Wignerdelay}
\end{figure}

The Wigner delay can be understood in a semi-classical model where a two-level atom elastically scatters a classical light field. In the limit of weak intensities, the atom responds linearly to the incoming light field, and the associated dispersive behaviour leads to a time delay of the re-emitted field that is maximum at resonance and is given by $\tau_{\rm W}~=~d\phi/d\omega$. Here, $\omega$ is the frequency of light, $\phi(\omega)=\arctan(\frac{\Gamma}{2(\omega_0-\omega)})$ is the phase of the atomic polarizability~\cite{Jackson,Allen}, $\omega_0=2\pi c/\lambda_{0}$ is the frequency of the atomic resonance, and $\Gamma$ is the inverse lifetime of the excited state $|e\rangle$ (see fig.~\ref{fig1_Wignerdelay}). $\tau_{\rm W}$ is thus given by
\begin{equation}\label{eq:formuletimedelay}
\tau_{\rm W}(\omega) = \frac{2}
{\Gamma}\, \frac{1}{1+4\left(\frac{\omega-\omega_{0}}{\Gamma}\right)^2}.
\end{equation}
When the incident wave is a weak pulse of light, the scattered field amplitude is obtained by adding coherently the scattered amplitudes associated to each Fourier component of the incident pulse, and $\tau_{\rm W}$ is the delay in the arrival time of the pulse envelope, induced by the presence of the atom~\cite{Wigner1955,Smith1960,Nussenzveig2002,Jackson}. In the particular case of a gaussian pulse of light with root-mean-square (rms) duration $\Delta t \gg 1/\Gamma$ and small intensity ($I/ I_{\rm sat}\ll 1$), the scattered pulse is undistorted, with an intensity given by
\begin{equation}\label{eq:formuleIntscat}
 I_{\rm sc}(r,t) \propto \frac{1}{1+4(\frac{\omega_{\rm L} -\omega_{0}}{\Gamma})^2} e^{-[t-\frac{r}{c}-\tau_{\rm W}(\omega_{\rm L})]^2/2\Delta t^2}.
\end{equation}
Here, $\omega_{\rm L}$ is the center frequency of the laser pulse spectrum, $t$ is time, $r$ is the distance between the observer and the scatterer, and $c$ is the velocity of light. In the limits mentioned above, we thus expect a scattered pulse that is gaussian in temporal shape and maximally delayed at resonance by $\tau_{\rm W}(\omega_0)=2/\Gamma$, which can reach several tens of nanoseconds for optical transitions of e.g. alkali atoms.

\begin{figure}
\centerline{\includegraphics[width=8cm]{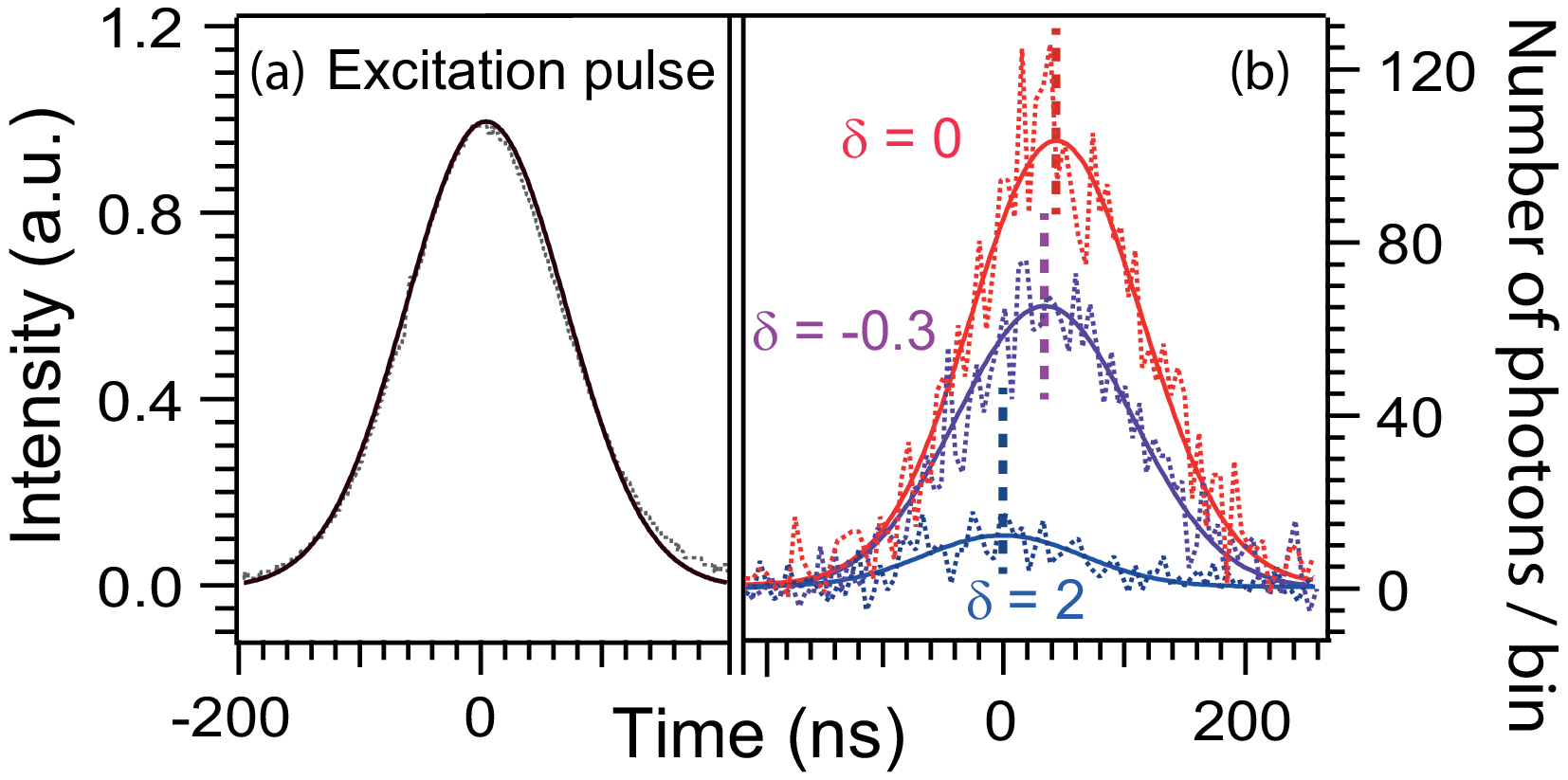}}
\caption{(a) Intensity of the excitation pulse with rms width $\Delta t=66$~ns. (b) Histograms of the number of photons detected after scattering for $\delta=(\omega_{\rm L}-\omega_{0})/\Gamma=(0; -0.3; 2)$ (time bins : $5.9$~ns). Dotted lines : measured data; solid lines : gaussian fits. In (b), the rms widths of the scattered pulses are, respectively, $(73; 73; 65)$~ns, in agreement with the solution of
eqs.~(\ref{EqBloch}).}\label{fig2_Wignerdelay}
\end{figure}

To check this prediction experimentally, we use a single cold $^{87}$Rb atom
that we initially isolate in an optical dipole trap with microscopic size~\cite{Sortais2007}. The temperature of the atom in the trap is $70~\mu$K,
measured by a release-and-recapture method~\cite{Tuchendler2008}.
We first prepare the trapped atom in the hyperfine ground state level $(5S_{1/2},F=2)$.
We then release it in free space and illuminate it with a series of weak gaussian
pulses of circularly polarized near-resonant laser light at $\lambda_{0}=780$~nm
(see fig.~\ref{fig1_Wignerdelay}). Using a large numerical aperture lens (N.A.=$0.5$),
we collect the photons scattered at $90^\circ$ with respect to the
direction of excitation and detect them with a fiber-coupled avalanche
photodiode (APD) operating in the single photon counting mode.
The signal is sent to a counting card
with a $256$~ps resolution. The excitation light is produced from a
continuous laser locked on the $(5S_{1/2}, F=2)$ to $(5P_{3/2},F'=3)$ transition.
We chop this light into pulses with a gaussian temporal shape using an acousto-optic
modulator and a fast arbitrary waveform generator. We set the peak intensity $I$
of the pulse to $I/ I_{\rm sat}=0.1$ ($I_{\rm sat}=1.6$~mW/${\rm cm}^2$) to
operate in the weak excitation limit. For the temporal width of the pulses,
we choose $\Delta t =66$~ns sufficiently large to approach the limit
$\Delta t\gg 1/\Gamma$ (here, $1/\Gamma=26$~ns), and sufficiently small
to determine with a good accuracy the temporal center of the scattered pulse,
and thus the time-delay $\tau_{\rm W}$ (see fig.~\ref{fig2_Wignerdelay}a).
The acousto-optic modulator allows us to tune the center frequency $\omega_{\rm L}$
of the laser light around the frequency $\omega_0$ of the closed transition
between the states $|g\rangle=|5S_{1/2},F=2,M_{\rm F}=2\rangle$ and
$|e\rangle=|5P_{3/2},F'=3,M'_{\rm F}=3\rangle$. The optical
pumping of the atom in the Zeeman sub-level $|g\rangle$
 is ensured by the  first excitation pulses when the laser is on resonance with the
 atom. When the laser is tuned away from the resonance,
 the optical pumping is less efficient. However the Wigner delay is unaffected
 by the events when the atom does not cycle on the closed transition, as both the
 center frequency and the width of the
 resonance, which sets the value of the delay, are independent of the transition between the states
 $|5S_{1/2},F=2,M_{\rm F}\rangle$ and  $|5P_{3/2},F'=3,M'_{\rm F}\rangle$.
 During the excitation, repumping
 light tuned to the $(5S_{1/2}, F=1)$ to $(5P_{3/2},F'=2)$ transition is
 also sent on the atom.

To maximize the number of collected photons scattered by the same atom
we use a time sequence where we interleave excitation pulses in free space
with recapturing periods of $1.3~\mu$s.
After $50$ such excitation-and-recapture periods, photon scattering has heated the atom and the
probability that the atom escapes the trap has increased. We  therefore apply a
$1$~ms period of three-dimensional laser-cooling with the dipole trap on.
In this way we keep the  Doppler shift below $100$~kHz.
We repeat this pattern $120$ times, corresponding to a total of $6000$ pulses
sent on the same atom before we start again with a newly prepared atom.

\begin{figure}
\centerline{\includegraphics[width=8cm]{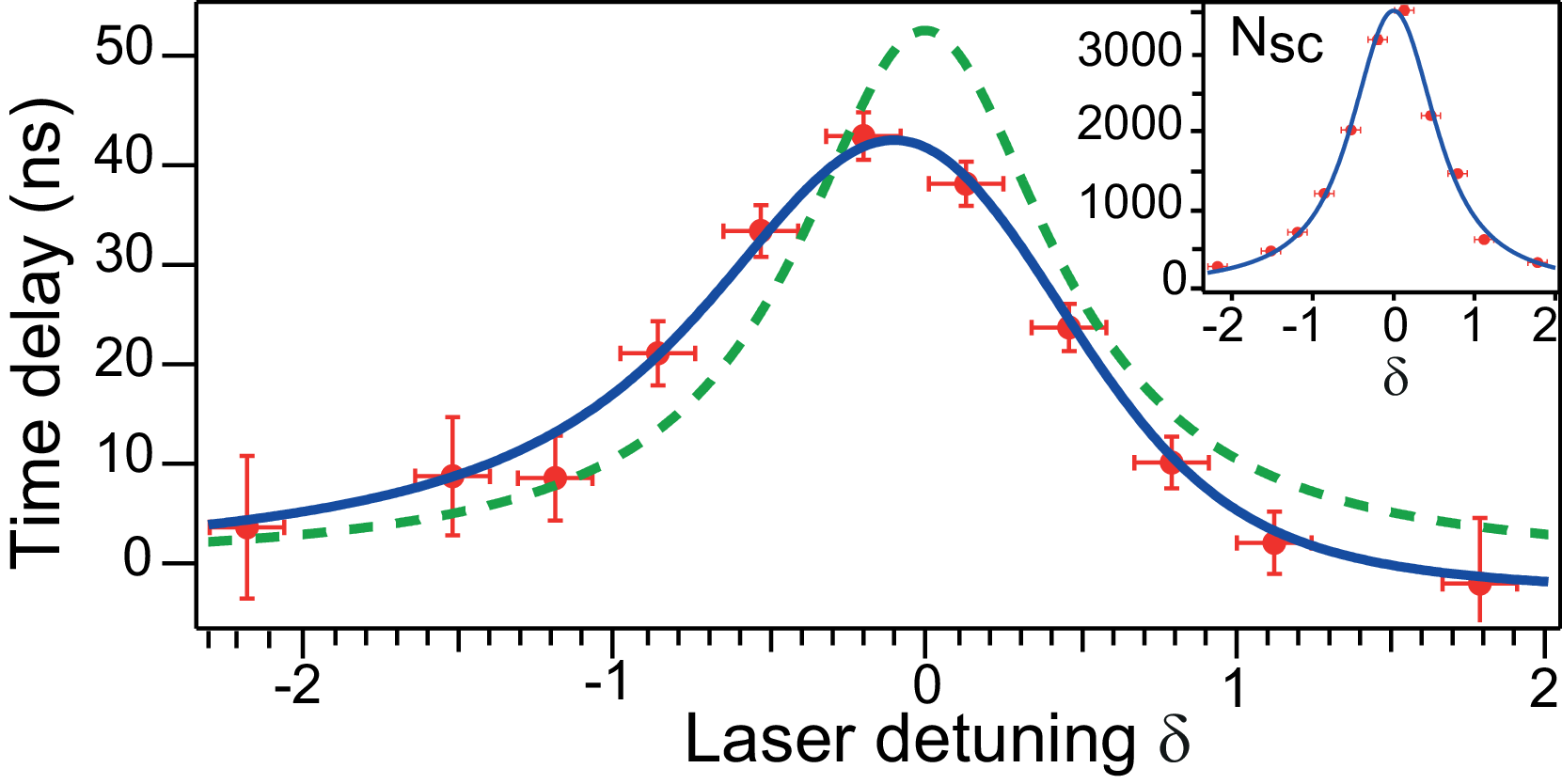}}
\caption{Time-delay of the scattered light pulse versus the detuning $\delta$ of the excitation laser. Dashed line: prediction of eq.~(\ref{eq:formuletimedelay}). Solid line : solution of eqs.~(\ref{EqBloch}) for a two-level atom excited by the pulse shown in fig.~\ref{fig2_Wignerdelay}a, with a fitted chirp rate $\alpha/2\pi=5\pm0.8$~MHz/$\mu$s. The vertical error bars are from the fits of the scattered pulses. Horizontal error bars: rms uncertainty on the laser frequency ($0.12\Gamma$). Inset: number of scattered photons detected on the APD. Solid line : solution of eqs.~(\ref{EqBloch}).}\label{fig3_Wignerdelay}
\end{figure}

Figure~\ref{fig2_Wignerdelay}b shows the temporal responses obtained
on the photon counter for different values of the detuning
$\delta=(\omega_{\rm L}-\omega_0)/\Gamma$ of the excitation laser. Each temporal
response results from an integration over $2000$ individual atoms having
experienced the sequence described above, and is fitted with a gaussian to
extract the arrival time of the scattered pulse.
At resonance, the scattered pulse is maximally delayed (and most intense),
as expected from eqs.~(\ref{eq:formuletimedelay}) and (\ref{eq:formuleIntscat}).
Figure~\ref{fig3_Wignerdelay} summarizes the variation of the arrival time
versus $\delta$. The comparison to far off-resonance
measurements reveals a Wigner time-delay $\tau_{\rm W}$ as large as $42\pm2$~ns
at resonance, not far from the $2/\Gamma=52$~ns predicted by eq.~(\ref{eq:formuletimedelay}).

The discrepancy is actually due to the conditions $\Delta t\gg1/\Gamma$
and $I/I_{\rm sat}\ll 1$ not being exactly fulfilled experimentally.
 In particular, the latter condition means that the scattering is not only elastic,
 as implicit from the model discussed above, but has also a small inelastic component.
 To take this into account, we solved the optical Bloch equations governing
 the evolution of the density matrix for a
 two-level atom, using the measured shape of the excitation
 pulse as a driving term.
 These equations  read in the rotating-wave approximation~\cite{Cohen}:
 \begin{eqnarray}\label{EqBloch}
 \dot{\tilde{\rho}}_{eg} &=& i (\omega_{\rm L}-\omega_0 (t))\, \tilde{\rho}_{eg} +i\frac{\Omega(t)}{2}\, (2\rho_{ee}-1)-\frac{\Gamma}{2}\tilde{\rho}_{eg}\nonumber\\
 \dot{\rho}_{ee} &=&i\frac{\Omega(t)}{2}\, (\tilde{\rho}_{eg}-\tilde{\rho}_{ge}) -\Gamma \rho_{ee}\ ,
 \end{eqnarray}
 where $\tilde{\rho}_{eg}=\rho_{eg}\, \exp{(i\omega_{\rm L}t)}$,
 $\tilde{\rho}_{eg}=\tilde{\rho}_{ge}^*$,
 $\Omega(t)=\Omega\, e^{-t^2/4\Delta t^2}$ and
 $2\Omega^2/\Gamma^2=I/I_{\rm sat}$. In order to account for the slight asymmetry observed on the data (see fig.~\ref{fig3_Wignerdelay}), we have allowed for a
 small chirp of the transition frequency during the pulse:
 $\omega_{0}(t)=\omega_0 + \alpha\, t$~\footnote{The exact origin of the chirp is still
 under investigation. It might come from a small variation of the residual light-shift
 of  the atomic transition frequency during the switching off of the dipole trap laser on this small
 time scale. }.
 The resolution of eqs.~\ref{EqBloch} yields the temporal evolution of the population
 $\rho_{ee}(t)$ in state $|e\rangle$, proportional to the scattered
 pulse intensity $I_{\rm sc}(t)$.
 For our experimental parameters $I_{\rm sc}(t)$
 is also gaussian to a very good
 approximation, which allows us to identify unambiguously the Wigner
 delay\footnote{We also note that the optical Bloch equations predict a
 temporal width of the scattered pulse that varies slightly with the detuning
 (see fig.~\ref{fig2_Wignerdelay}b).}. A fit of the data by this second model
with the chirp rate $\alpha $ as an adjustable parameter
 gives $\alpha/2\pi=5\pm 0.8$~MHz/$\mu$s and  predicts a maximum delay of $42$~ns at resonance, in very good agreement with the measured delay. By setting $I/I_{\rm sat} < 0.01$ and
 $\Delta t > 10/\Gamma$ in eqs.~(\ref{EqBloch}), as well as $\alpha=0$, we checked that
 we recovered the asymptotic value of the Wigner delay
 $2/\Gamma$ characteristic of the  elastic scattering regime to better than 2\%.

To characterize the scattering process fully we also analyze the number of
collected photons as a function of the excitation detuning $\delta$.
To do so, we integrate the temporal signal obtained on the APD. The data are again in good agreement with the solution of eqs.~(\ref{EqBloch}), and exhibit a full-width at half maximum of
$1.2\Gamma$ (see fig.~\ref{fig3_Wignerdelay} inset),
larger than the lorentzian profile predicted by eq.~(\ref{eq:formuleIntscat}).
This larger width is also consistent with the reduced time-delay due to the
conditions $\Delta t\gg1/\Gamma$
and $I/I_{\rm sat}\ll 1$ not being exactly fulfilled experimentally.

In conclusion, we have measured the time-delay introduced by an
individual atom in an essentially elastic scattering process.
We have found delays as large as $42$~ns in good agreement
with the theoretical limit predicted by the optical Bloch equations.
In the future it will be interesting to extend these measurements to
the case of dense ensembles of cold interacting atoms to probe
collective scattering~\cite{Lagendijk1996}.\\

We acknowledge support from the E.U. through the ERC Starting Grant
ARENA and the Integrated Project AQUTE, and from the Triangle de la Physique.
We thank G.~Labeyrie, T.~Lahaye, A.~Vernier and
A.~Aspect for useful discussions.

%\pagebreak


\begin{thebibliography}{99}
\bibitem{Wigner1955} E.P.~Wigner, "Lower limit for the energy derivative of the scattering phase shift", Phys. Rev. {\bf 98}, 145 (1955).

\bibitem{Smith1960} F.T.~Smith, "Lifetime matrix in collision theory", Phys. Rev. {\bf 118}, 349 (1960).

\bibitem{Lagendijk1996} A.~Lagendijk and B.A.~van~Tiggelen, "Resonant multiple scattering of light", Physics Reports {\bf 270}, 143 (1996).

\bibitem{Nussenzveig2002} C.A.A.~de~Carvalho and H.M.~Nussenzveig, "Time delay", Physics Reports {\bf 364}, 83 (2002).

\bibitem{Labeyrie2003} G.~Labeyrie, E.~Vaujour, C.A.~M\"uller, D.~Delande, C.~Miniatura, D.~Wilkowski and R.~Kaiser, "Slow diffusion of light in a cold atomic cloud", Phys. Rev. Lett. {\bf 91}, 223904 (2003).

\bibitem{MullerPRA2005} C.A.~M\"uller, C.~Miniatura, D.~Wilkowski, R.~Kaiser and D.~Delande, "Multiple scattering of photons by atomic hyperfine multiplets", Phys. Rev. A \textbf{72}, 053405 (2005).

%\bibitem{PierratPRA2009} R.~Pierrat, B.~Gr\'emaud and D.~Delande, "Enhancement of radiation trapping for quasiresonant scatterers at low temperature", Phys. Rev. A \textbf{80}, 013831 (2009).

\bibitem{PierratPRA2010} R.~Pierrat and R.~Carminati, "Spontaneous decay rate of a dipole emitter in a strongly scattering disordered environment", Phys. Rev. A \textbf{81}, 063802 (2010).

\bibitem{Maroni1976} C.~Maroni, I.~Massa and G.~Vannini, "Time delay measurements in a low-energy nuclear reaction from a bremsstrahlung experiment", Phys. Lett. {\bf 60} B, 344 (1976).

\bibitem{Chauvat2000} D.~Chauvat, O.~Emile, F.~Bretenaker and A.~Le~Floch,
"Direct measurement of the Wigner delay associated with the Goos-Hänchen effect",
Phys. Rev. Lett. {\bf 84}, 71 (2000).

\bibitem{Cavalieri2007} A.L.~Cavalieri, N.~M\"uller, Th.~Uphues, V.S.~Yakovlev, A.~Baltu\u{s}ka, B.~Horvath, B.~Schmidt, L.~Bl\"umel, R.~Holzwarth, S.~Hendel, M.~Drescher, U.~Kleineberg, P.M.~Echenique, R.~Kienberger, F.~Krausz and U.~Heinzmann,
"Attosecond spectroscopy in condensed matter",
Science \textbf{449}, 1029 (2007).

\bibitem{Schultze2010} M.~Schultze, M.~Fie{\ss}, N.~Karpowicz, J.~Gagnon, M.~Korbman, M.~Hofstetter, S.~Neppl, A.L.~Cavalieri, Y.~Komninos, Th.~Mercouris, C.A.~Nicolaides, R.~Pazourek, S.~Nagele, J.~Feist, J.~Burgd\"orfer, A.M.~Azzeer, R.~Ernstorfer, R.~Kienberger, U.~Kleineberg, E.~Goulielmakis, F.~Krausz and V.S.~Yakovlev,
"Delay in photoemission",
Science \textbf{328}, 1658 (2010).

\bibitem{Klunder2011} K.~Kl\"under, J.M.~Dahlstr\"om, M.~Gisselbrecht, T.~Fordell, M.~Swoboda, D.~Gu\'enot, P.~Johnsson, J.~Caillat,
J.~Mauritsson, A.~Maquet, R.~Ta\"ieb and A.~L'Huillier,
"Probing single-photon ionization on the attosecond time scale",
Phys. Rev. Lett. {\bf 106}, 143002 (2011).

\bibitem{Jackson} J.D.~Jackson, ``Classical Electrodynamics'', Wiley (1998).

\bibitem{Allen} L.~Allen and J.H.~Eberly, ``Optical resonance and two-level atoms'', Dover (1987).

\bibitem{Sortais2007} Y.R.P.~Sortais, H.~Marion, C.~Tuchendler, A.M.~Lance, M.~Lamare, P.~Fournet, C.~Armellin, R.~Mercier, G.~Messin, A.~Browaeys and
P.~Grangier,
"Diffraction-limited optics for single atom manipulation",
Phys. Rev. A \textbf{75}, 013406 (2007).

%\bibitem{Schlosser2001} N.~Schlosser, G.~Reymond, I.~Protsenko and P.~Grangier,
%"Sub-poissonian loading of single atoms in a microscopic dipole trap",
%Nature (London) {\bf 411}, 1024 (2001).

\bibitem{Tuchendler2008} C.~Tuchendler, A.M.~Lance, A.~Browaeys, Y.R.P.~Sortais, and P.~Grangier,
"Energy distribution and cooling of a single atom in an optical tweezer",
Phys. Rev. A \textbf{78}, 033425 (2008).

%\bibitem{Darquie2005} B.~Darqui\'e, M.P.A.~Jones, J.~Dingjan, J.~Beugnon, S.~Bergamini, Y.~Sortais, G.~Messin, A.~Browaeys and P.~Grangier,
%"Controlled single-photon emission from a single trapped two-level atom", Science {\bf 309}, 454 (2005).

\bibitem{Cohen} C. Cohen-Tannoudji, J. Dupont-Roc, and G. Grynberg, ``Photons and atoms: basic
process and applications'', Wiley (1997).

\end{thebibliography}
\end{document}